\documentclass[10pt,journal,compsoc]{IEEEtran}

\ifCLASSOPTIONcompsoc
  \usepackage[nocompress]{cite}
\else
  \usepackage{cite}
\fi

\ifCLASSINFOpdf

\else

\fi

\usepackage{graphicx}
\usepackage{tabularx}
\usepackage{amsmath}
\usepackage{cite}
\usepackage[square, comma, sort&compress, numbers]{natbib}
\begin{document}

\title{Multi-Wavelet Residual Dense Convolutional\\ Neural Network for Image Denoising}

\iftrue
\author{Shuo-Fei~Wang,
        ~Wen-Kai~Yu,
        ~and~Ya-Xin~Li
\IEEEcompsocitemizethanks{\IEEEcompsocthanksitem S.-F Wang, W.-K. Yu, and Ya-Xin Li are with Center for Quantum Technology Research, and Key Laboratory of Advanced Optoelectronic Quantum Architecture and Measurement of Ministry of Education, School of Physics, Beijing Institute of Technology,
Beijing 100081, China.\protect\\
E-mail: yuwenkai@bit.edu.cn
\IEEEcompsocthanksitem S.-F Wang and W.-K. Yu contributed equally to this work.}
\thanks{Manuscript received XX XX, 2020; revised XX XX, 2020.\protect\\
(Corresponding author: Wen-Kai Yu.)}}

\fi
\IEEEtitleabstractindextext{%

\begin{abstract}
Networks with large receptive field (RF) have shown advanced fitting ability in recent years. In this work, we utilize the short-term residual learning method to improve the performance and robustness of networks for image denoising tasks. Here, we choose a multi-wavelet convolutional neural network (MWCNN), one of the state-of-art networks with large RF, as the backbone, and insert residual dense blocks (RDBs) in its each layer. We call this scheme multi-wavelet residual dense convolutional neural network (MWRDCNN). Compared with other RDB-based networks, it can extract more features of the object from adjacent layers, preserve the large RF, and boost the computing efficiency. Meanwhile, this approach also provides a possibility of absorbing advantages of multiple architectures in a single network without conflicts. The performance of the proposed method has been demonstrated in extensive experiments with a comparison with existing techniques.
\end{abstract}

}

\maketitle

\IEEEdisplaynontitleabstractindextext

\IEEEpeerreviewmaketitle

\IEEEraisesectionheading{\section{Introduction}}

\IEEEPARstart{I}{mage} denoising is one of well-known ill-posed problems. The noisy images are typically caused by external noise like electromagnet wave interruption \cite{Matter-TMI-2008} or internal noise from the detectors themselves \cite{Kim-EL-2017}. To meet the needs of extensive applications, many image denoising algorithms have been proposed for decades \cite{Gilboa-TPAMI-2004,Coupe-TMI-2008,Luisier-TIP-2008,Montagner-TIP-2014,Dabov-TIP-2007}. Recently, benefiting from the increasingly computing power brought by advanced graphic processing units (GPU), various neural networks, especially convolutional neural network (CNN), have been raised up to resolve image denoising tasks. Deserved to be mentioned, in CNN, once the network is trained, the image denoising procedure can be accomplished with a simple forward propagation, which will significantly reduce the computing time, while the degraded image can be treated as a non-linear map from the clean image \cite{PLiu-IA-2019}. Due to the employment of the nonlinear activation function, the CNN are gifted in dealing with noisy image denoising tasks.

During this decade, the CNN has shown a more effective performance in comparison to traditional methods. Although there are a variety of networks, most of them can be divided into three categories in terms of their structure features, including simple, multi-residual and U-net structures. The networks with simple structure do not have skip connection, which means that the feature maps propagate layer-by-layer. Due to the low computing load and convenient designing process, CNNs with simple structure are widely used in the early attempts \cite{Venetianer-TCS-1995,Zeng-TMI-2014,Zineddin-TIP-2011}. However, the connection between each convolution layers is neglected in these networks, and dead neurons will be incurred due to the gradients vanishing problem in very deep networks \cite{He-CVPR-2016}. In 2015, He et al. realized residual learning by inserting shortcut connections, which perform element-wise addition between inputs and outputs \cite{He-CVPR-2016}. This multi-residual structure is helpful for maintaining suitable gradients in back propagation and providing the possibility of obtaining a better performance in deep networks. Since then, several networks with more complicated connections were presented to take full advantage of the relationship between each layer \cite{CLiu-IA-2019,Wei-EL-2019,Mao-ANIPS-2016}. The U-net structure was first proposed by Ronneberger et al. to apply CNN for biological segmentation \cite{Ronneberger-MICCAI-2015}. U-net gets its name from its characteristic architecture whose backbone is composed of pooling and up-convolution layers. Due to the use of the pooling layer, U-net significantly reduces the calculation cost. Besides, U-net contains element-wise addition or concatenation connections to establish links between the layers in the same level. Recently, Liu et al. \cite{PLiu-IA-2019} proposed to adopt discrete wavelet transform (DWT) and inverse discrete wavelet transform (iDWT) in U-net to eliminate the information loss caused by the pooling operation . Their multi-level wavelet convolutional neural network (MWCNN) \cite{PLiu-IA-2019} further enhanced the precision of the deep learning technique in image denoising tasks.

\begin{figure}[!t]
\centering
\includegraphics[width=3.5in]{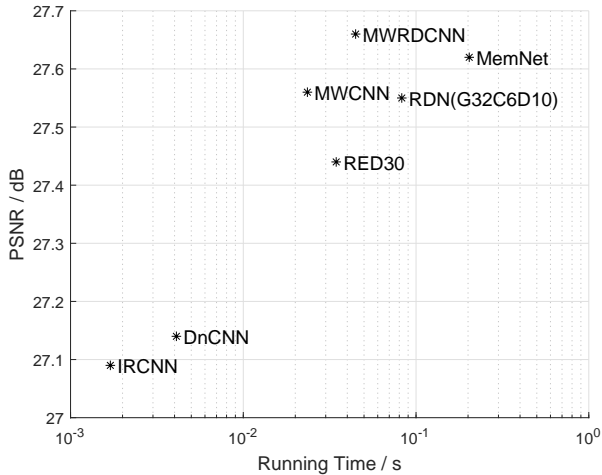}
\caption{The denoising performance of our MWRDCNN and six state-of-the-art networks on training sets (Set5 was used here) with the standard deviation of the additive Gaussian noise $\sigma=50$. The average PSNRs and running time of the same single test image (of $0\sim255$ gray-scale) were evaluated with a RTX 2080Ti GPU.}
\label{fig:PSNRtime}
\end{figure}

From a certain perspective, multi-residual and U-net structures are the optimized variants of simple structure.
Here, we chose six representative networks as references, including simple structures: learning deep CNN denoiser prior for image restoration (IRCNN) \cite{Zhang-CVPR-2017}, denoising convolutional neural network (DnCNN) \cite{Zhang-TIP-2017}; multi-residual structures: residual encoder-decoder network with 30 layers (RED30) \cite{Mao-ANIPS-2016}, residual dense network (RDN) with 32 feature map channels before entering 6 RDBs each of which contains 10 convolutional layers (G32C6D10) \cite{Zhang-CVPR-2018}, memory network (MemNet) \cite{Tai-ICCV-2017}; and U-net structure: MWCNN \cite{PLiu-IA-2019}. As shown in Fig.~\ref{fig:PSNRtime}, both multi-residual and U-net structures lead to a higher peak signal-to-noise ratio (PSNR) than the simple structure. Nevertheless, both multi-residual and U-net structures have disadvantages: the former takes enormous computation time, the latter has fewer internal connections and the residual learning is less efficient. Therefore, to make complementary advantages of these two kinds of networks, we propose a more robust CNN by combining both the MWCNN structure and residual dense blocks (RDBs) together, i.e., it can be categorized into U-net \& multi-residual structure. We refer to this novel network as multi-wavelet residual dense convolutional network (MWRDCNN).

We noticed that the networks reported in different papers were generally trained with different training sets and different parameters, which will definitely affect the final results \cite{Agustsson-CVPRW-2017}. Thus, to make a fair comparison, we trained these representative networks in the same condition with our network.

The main contributions of this work include:
\begin{itemize}
\item adopting the RDBs in our MWCNN architecture to obtain a better tradeoff between the computation speed and denoising performance;
\item validating the effectiveness of RDBs in image denoising tasks;
\item demonstrating the performance of our MWCNN architecture compared with six state-of-the-art networks under the same training conditions.
\end{itemize}

\section{Related Works}


\subsection{Traditional Denoising Algorithms and Early CNN Structures for Image Denoising}
It is worth noting that before the CNN was applied in image denoising tasks, the conventional methods were relatively mature \cite{Wang-TIP-2006,Lian-TIP-2006,Hirakawa-TIP-2006,Bertalmio-TIP-2007,Kervrann-TIP-2006}. In 2007, Dabov et al. proposed block-matching and 3D filtering (BM3D) method \cite{Dabov-TIP-2007}, which was once thought to be the most efficient traditional denoising algorithm. So in the early years when the CNN was adopted to solve the ill-posed problems of image denoising, the BM3D method became the reference standard of the performance evaluation of newly proposed methods. Jain and Seung proposed a four-layer convolutional network \cite{Jain-ANIPS-2008}, which was proved to have a superior performance compared with the conventional wavelet and Markov random field (MRF) methods \cite{Lan-CV-2006}. Xie et al. presented a deep network scheme utilizing stacked sparse denoising auto-encoders \cite{Xie-ANIPS-2012}, which achieved a compatible performance against K-singular value decomposition (K-SVD) algorithm \cite{Elad-TIP-2006}. Although those CNN-based methods transcended the most conventional algorithms, they failed to achieve better results compared with the BM3D method. The monopoly of BM3D in single image super resolution (SISR) was broken in 2014 by a three-layer fully convolutional network (FCN) proposed by Dong et al. \cite{Dong-TPAMI-2016}. Then, Zhang et al. enlarged the depth of the network to 17 layers \cite{Zhang-TIP-2017}, significantly boosting its performance. Meanwhile, it was the first time that the CNN was used for image denoising. Since then, many kinds of convolution networks with very deep architecture have been developed \cite{Zhang-IA-2019,Hashimoto-IA-2019,Shan-IA-2019}.

\subsection{Residual Learning}
A typical method to conduct residual learning is to adopt addition operation at the end of the network, which is promised to solve the gradient vanishing problem to a certain extent. Nevertheless, as the depth of the network grows, it is harder for very deep networks to keep a long-term memory. Some intuitive solutions are to establish more complicated connections among the convolutional layers. For instance, Mao et al. \cite{Mao-ANIPS-2016} considered to establish multiple connections between encoders and decoders by element-wise addition; Tai et al. \cite{Tai-ICCV-2017} adopted the dense connected memory blocks to take into account both short-term memory and long-term memory; Zhang et al. further proposed the residual dense block to achieve full use of all convolutional layers \cite{Zhang-CVPR-2018}.

\subsection{U-net Architecture and DWT in CNNs}
To avoid overfitting and to reduce the amount of calculation, pooling layers are commonly employed in CNNs. This means that the sizes of feature maps reduce along the forward propagation. Thus, early CNNs are often used to solve classification problems by giving a single label. However, the objectives of image processing like image segmentation and denoising are not confined to classification problems, and they require the outputs to be the images that have almost the same size as the inputs. In 2015, Ronneberger et al. adopted the up-convolutional layers, which enlarged the sizes of feature maps by convolution and expanded the application of CNN to biomedical segmentation \cite{Ronneberger-MICCAI-2015}. Since this network owns a novel architecture, they called this network U-net. In U-net, the calculations are accelerated, and the pooling layer is a convenient tool to enlarge the receptive field (RF). Therefore, the CNNs with U-net structure seem to be more suitable to deal with image denoising tasks compared with the traditional networks. But, the pooling operation can also cause inevitable information loss. Inspired by conventional wavelet algorithm, Bae et al. put forward a wavelet residual network (WavResNet) by adopting DWT and iDWT layer instead of pooling layer \cite{Bae-CVPRW-2017}. Later, Liu et al. optimized this network on the basis of U-net and proposed the MWCNN \cite{PLiu-IA-2019}, which improved the image denoising performance.

The aforementioned works have succeeded in comprehensive learning, calculation simplifying and RF enlarging. To absorb the advantages of those networks, here we propose a multi-level wavelet residual dense convolutional neural network. The architecture of the network will be detailed in the next section.

\section{Proposed Method}
In this section, we will first briefly review the procedures of DWT and iDWT to introduce the foundation of our proposed method. Then, we will formally show our MWRDCNN based on DWT, iDWT and RDBs. The details of RDBs will be also introduced. After that, the implementation details and the difference from the previous works will be presented.

\subsection{DWT and iDWT}
\begin{figure}[ht]
\centering
\includegraphics[width=3.5in]{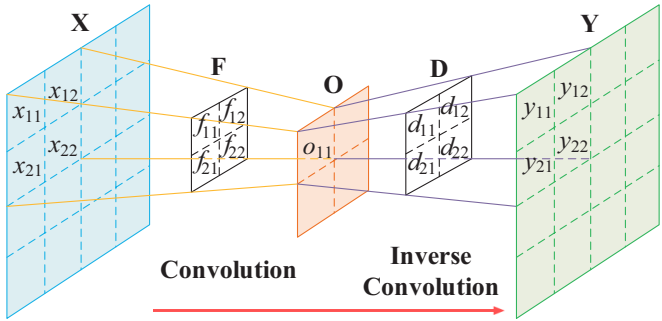}
\caption{Schematic diagram of convolution and inverse convolution processes.}
\label{fig:inverse}
\end{figure}

Before introducing the DWT and iDWT, we first draw the schematic diagram of convolution and inverse convolution processes in CNNs, as shown in Fig.~\ref{fig:inverse}. Taking the four pixels in the top left corner of the input image $\mathbf{X}$ as an example, when they are convolved with a $2\times2$ filter $\mathbf{F}$, we will have the first pixel value in the hidden layer $\mathbf{O}$ via
\begin{equation}
o_{11}=\sum_{i=1}^{2}\sum_{j=1}^{2}x_{ij}\times f_{ij}.
\end{equation}
In the direction of forward propagation, the pixel values of unknown $\mathbf{Y}$ will be obtained by
\begin{equation}
y_{ij}=o_{11}\times d_{ij},\ i,j\in[1,2].
\end{equation}
By performing the same operation for the rest pixels with a stride of 2, we can get the output image $\mathbf{Y}$ via inverse convolution between $\mathbf{O}$ and a $2\times2$ inverse filter $\mathbf{D}$:
\begin{equation}
\mathbf{Y}=Iconv(\mathbf{O},\mathbf{D}),
\end{equation}
where $Iconv$ represents the inverse convolution operation.

Actually, from the hidden layer's point of view, $o_{11}$ can also be symmetrically viewed as the convolution result of the four pixels in the top left corner of the output image $\mathbf{Y}$ and the dual filter $\mathbf{D}^\prime$ of $\mathbf{D}$, as illustrated in Fig.~\ref{fig:inverse}.

Now, let us review the concepts of the DWT and iDWT, which are the common tools in conventional image processing. A given image $\mathbf{X}$ can be decomposed into four sub-images by DWT, i.e. low-pass image $\mathbf{X}_{\mathrm{A}}$ (average), and three high-pass images including $\mathbf{X}_{\mathrm{H}}$ (horizontal), $\mathbf{X}_{\mathrm{V}}$ (vertical), $\mathbf{X}_{\mathrm{D}}$ (diagonal). From a certain perspective, the process of the DWT can seen as a convolution between $\mathbf{x}$ and four $2\times2$ filters described as below, in a stride of 2:
\begin{equation}
\begin{aligned}
\begin{array}{c}
\mathbf{f}_\mathrm{A}=\left[{\begin{array}{*{20}{c}}
\ \ 1\ &\ \ \ 1\\
\ \ 1\ &\ \ \ 1
\end{array}}\right],\quad\mathbf{f}_\mathrm{H}=\left[{\begin{array}{*{20}{c}}
-1&\ \ \ 1\\
-1&\ \ \ 1
\end{array}}\right],\vspace{0.2cm}\\
\mathbf{f}_\mathrm{V}=\left[{\begin{array}{*{20}{c}}
-1&-1\\
\ \ 1&\ \ \ 1
\end{array}}\right],\quad\mathbf{f}_\mathrm{D}=\left[{\begin{array}{*{20}{c}}
\ \ 1&-1\\
-1&\ \ \ 1
\end{array}}\right].
\end{array}
\end{aligned}
\end{equation}
Thus, through the DWT process, we will obtain four decomposed sub-images via
\begin{equation}
\mathbf{X}_i=Conv\left(\mathbf{X},\mathbf{f}_{i}\right),\ i=\mathrm{A},\mathrm{H},\mathrm{V}\text{ or }\mathrm{D},
\end{equation}
where $Conv$ refers to the convolution operation, $\mathbf{X}$ can be treated as the feature maps, and $\mathbf{f}_{i}$ denote the filters. As a result, the pixel-sizes of the four sub-images are half of the input image size. To some extent, the DWT has the similar downsampling effect as the pooling operation in U-net \cite{Ronneberger-MICCAI-2015}. Moreover, due to the orthogonality of the four filters, there will be no information loss during the downsampling process, which means that the target image can be completely reconstructed by iDWT. Previously, Liu et al. \cite{PLiu-IA-2019} adopted four equations to describe the iDWT process. Here, we find iDWT can also be expressed as an inverse convolution operation
\begin{equation}
\mathbf{\tilde{X}}=Sum\left(Iconv\left(\mathbf{X}_i,\mathbf{f}_i/ 4\right)\right),\ i=\mathrm{A},\mathrm{H},\mathrm{V}\text{ and }\mathrm{D},
\end{equation}
where $Sum$ denotes the element-wise addition.

Due to the information lossless property, the DWT and iDWT are gradually utilized in CNNs \cite{Guo-CVPRW-2017}. Among them, the MWCNN is one of the most representative networks \cite{PLiu-IA-2019}. Thus, here we adopt the MWCNN as the main framework of our work. The channel number of feature maps is set to $c$. Assuming the sizes of the feature maps before the DWT are all $h\times w$, then the sizes of the decomposed images are all $\frac{h}{2}\times\frac{w}{2}$. Several pairs of DWT and iDWT operations are sequentially used to establish our hierarchical structure, which will be introduced later.

\subsection{Network Architecture}
\begin{figure*}[ht]
\centering
\includegraphics[scale=0.6]{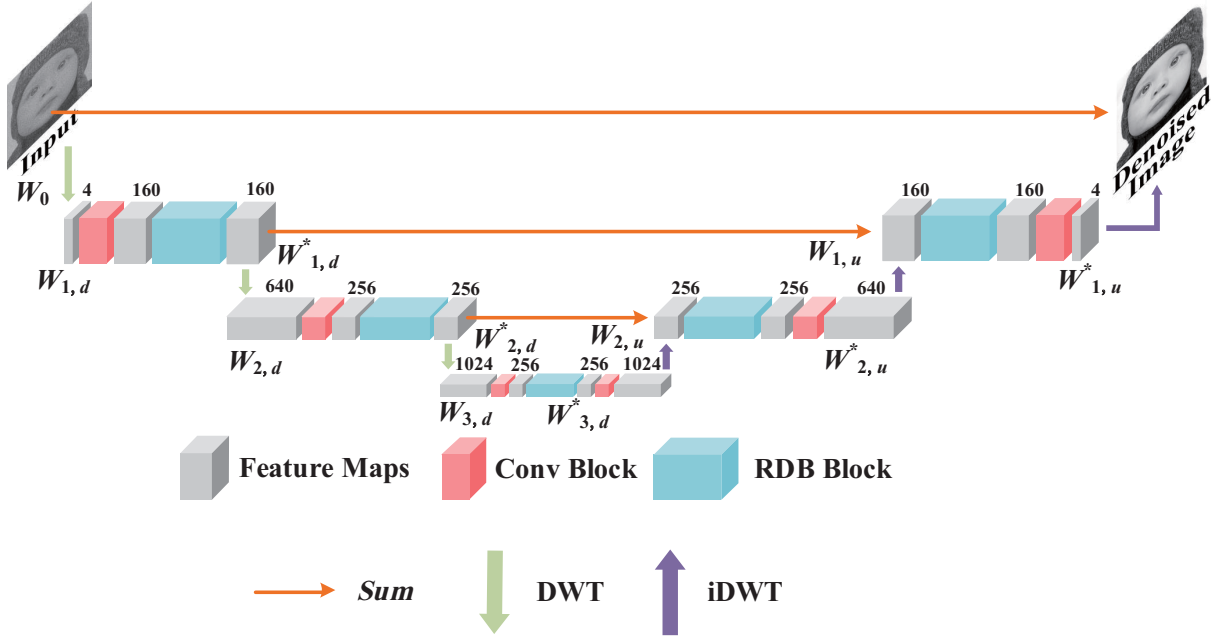}
\caption{MWRDCNN architecture. The number of channels is annotated on the top of the multi-channel feature map box. The names of the mentioned representative feature maps are also marked under the boxes.}
\label{fig:network}
\end{figure*}

As shown in Fig.~\ref{fig:network}, we retain the backbone of the MWCNN and optimize each level with the RDB to build the our network. Let us consider a three-level MWRDCNN. We denote the input image as $W_0$. The first DWT decompose $W_0$ into four subband feature maps, all marked as $W_{1}$. For the reason that these feature maps locate in the downsampling procedure, we further refer to them as $W_{1,d}$. The subband images in the upsampling procedure corresponding to $W_{1,d}$ will all be denoted by $W_{1,u}$. For the first DWT, we have
\begin{equation}
W_{1,d}=DWT(W_0).
\end{equation}
After that, a single convolution block (i.e., the Conv block in Fig.~\ref{fig:network}) is deployed to reduce the channels of the feature maps for improving inter-band independency of feature maps \cite{Guo-CVPRW-2017} and decreasing computation time cost. Whether the block contains a batch norm layer depends on the number of the current layer. Then, after the subband images pass through the RDB, we will obtain $W_{1,d}^*$ via
\begin{equation}
W_{1,d}^*=RDB(CB(W_{1,d})),
\end{equation}
where $CB$ denotes the function of the convolution block. After accomplishing local feature fusion by RDB, the second DWT is adopted to produce a hierarchical architecture in the network, where the subband images in the second level are expressed as
\begin{equation}
W_{2,d}=DWT(W_{1,d}^*).
\end{equation}

According to this procedure, the feature maps in the $i$th level during the downsampling procedure can be deduced from the equations below
\begin{equation}\label{eq:WWstar}
\left\{\begin{array}{c}
W_{i,d}=DWT(W_{i-1,d}^*),\\
W_{i,d}^*=RDB(CB(W_{i,d})).
\end{array}\right.
\end{equation}

Suppose that the network contains $n$ levels, the upsampling process begins after the feature maps passing the $n$th RDB. To guarantee the symmetric of this architecture, another convolution block will be used before the first iDWT operation. The iDWT in the highest level can be formally described as
\begin{equation}
W_{n-1,u}=iDWT(CB(W_{n,d}^*)),
\end{equation}
where $W_{n-1,u}$ refers to the $(n-1)$th upsampled feature maps. Then, the element-wise addition is adopted to establish long-term residual learning between the feature maps $W_{n-1,u}$ and $W_{n-1,u}^*$ in the same level. In the downsampling procedure, each layer is preceded by the convolution block, then followed by the RDB; while during the upsampling process, the order of the convolution block and RDB in each layer needs to be reversed, i.e., set the RDB first, then deploy the convolution block. As mentioned earlier, the role of the convolution blocks in the downsampling process is to reduce the number of channels, the ones in the upsampling procedure have precisely the opposite effect, i.e., to increase the number of channels and make it equal to the channel number of corresponding feature maps in the downsampling procedure of the same level, for the next element-wise addition. Subsequently, we will obtain the output feature maps $W_{n-1,u}^*$ by
\begin{equation}
W_{n-1,u}^*=CB(RDB(Sum(W_{n-1,u},W_{n-1,u}^*))).
\end{equation}
Similar to Eq.~(\ref{eq:WWstar}), the general term formula in the upsampling can be expressed as
\begin{equation}
\left\{\begin{array}{c}
W_{i-1,u}=iDWT(W_{i,u}^*),\\
W_{i-1,u}^*=CB(RDB(Sum(W_{i-1,u},W_{i-1,d}^*))).
\end{array}\right.
\end{equation}
It is worth noting that there are no more learnable parameters after the last convolution layer, so adding one more rectified linear unit (ReLU) excitation layer will definitely degrade the quality of the final output image. Therefore, we set the last convolution block be a single convolution layer.

In this work, we use the PSNR as a unitless performance measure for evaluating the quality of the final output images from the networks:
\begin{equation}
PSNR=10\log_{10}\left(\frac{255^{2}}{MSE}\right).
\end{equation}
From above definition we can see that the PSNR value depends on the mean square error (MSE), which is defined as $\textrm{MSE}=\frac{1}{pq}\sum\nolimits_{i,j=1}^{p,q}[U_o(i,j)-\tilde U(i,j)]^2$, where $p$ and $q$ are the pixel size of the input image and output image. The MSE can describe the squared distance between the denoised image $\tilde U(i,j)$ and the original image $U_o(i,j)$. Naturally, the larger is the PSNR value, the better is the image quality of the output result.

Since the quadratic term will come out a 2 during the derivative of backpropagation when training the network, to cancel out this 2, we define the loss function of our network as:
\begin{equation}
L_2(P)=\frac{1}{2N}\sum_{i=1}^N\left\|\Psi\left(\mathbf{b}_i,P\right)-\mathbf{x}_i\right\|_2^2,
\end{equation}
where $\mathbf{b}_i$ and $\mathbf{x}_i$ are the image with additive noise (the input image of the network) and the original image (the target, also called the ground truth), respectively, both of them comprise a training set $\left\{(\mathbf{b}_i,\mathbf{x}_i)\right\}_{i=1}^N$, $N$ is the batch size, $\Psi\left(\mathbf{b}_i,P\right)$ refers to the output image of the network, and $P$ denotes the learnable parameters in the network.

\subsection{Residual Dense Block}
\begin{figure}[ht]
\centering
\includegraphics[width=3.5in]{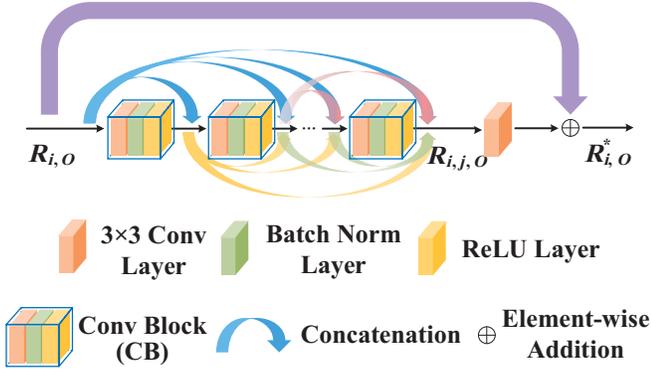}
\caption{Structure of residual dense blocks in our MWRDCNN.}
\label{fig:RDB2}
\end{figure}
The RDB is a common structure to build tight connections among layers and has been extensively used to deal with the SISR tasks \cite{Zhang-CVPR-2018}. In this work, we innovatively apply the RDBs to image denoising. The detailed architecture of RDBs is given in Fig.~\ref{fig:RDB2}. Assume that the inputs and outputs of the RDB in $i$th level are denoted as $R_{i,o}$ and $R_{i,o}^*$, respectively, where $o=d$ (downsampling) or $u$ (upsampling), we can simply write the relationship as below:
\begin{equation}
R_{i,o}^*=RDB(R_{i,o}).
\end{equation}
For a RDB with $j$ convolution blocks, the output of the final concatenation result $R_{i,j,o}$ can be derived from
\begin{equation}
R_{i,j,o}=Concat(CB(R_{i,j-1,o}),R_{i,j-1,o},\cdots,R_{i,2,o},R_{i,o}),
\end{equation}
where $Concat$ denotes the concatenation operation. Assuming the number of channels in $R_{i,o}$ is $c$, the $R_{i,j,o}$ will contain $j\times c$ channels. Thus, another convolution layer is needed to fuse the feature maps for the local residual learning. Then, we compute $R_{i,o}^*$ via
\begin{equation}
R_{i,o}^*=Sum(R_{i,o},Conv(R_{i,j,o})).
\end{equation}

\subsection{Implementation Details}
Our MWRDCNN consists of three hierarchical levels. Each RDB contains three convolution blocks. The sizes of the convolution kernels in the convolution layers are all set to $3\times3$, and the weights of each convolution kernel need to be adjusted in backpropagation; the sizes of the convolution kernels in the DWT and iDWT are all $2\times2$, and instead their weights are not adjusted in backpropagation. It is worth mentioning that there are no convolution layers in the DWT and iDWT. The numbers of channels in each layer are marked in Fig.~\ref{fig:network}.

\subsection{Discussion}
By combining with the preeminent structures of the RDN and MWCNN, we will discuss our efforts to address the trade-off between performance and time consumption by using our MWRDCNN.

\subsubsection{Difference between the MWRDCNN and RDN}
Deserved to be mentioned, the RDN does not contain a U-shape structure, so only two convolution layers are taken into account in the RDN to extract the shallow features. In our network, the subband images are the new feature maps decomposed from the input images, which means that their features are needed to be extracted. Therefore, we add one more convolution block in each level of the network.

\subsubsection{Difference between the MWRDCNN and MWCNN}
Here, we summarize three main differences between our network and the MWCNN. First, in the MWCNN, only long-term residual learning is applied by element-wise adding the downsampling and upsampling feature maps in the same level. Since we adopt the RDBs in each layer of our MWRDCNN, both the short-term and long-term connections are established, resulting in a more efficient learning mechanism. Second, in our MWDDCNN, we rewrite the DWT and iDWT functions into convolution and inverse convolution operations, instead of matrix operations used in MWCNN. Third, each RDB in our MWRDCNN contains 3 convolution blocks, so the downsampling or upsampling process of the first two layers separately contains 4 convolution blocks, which is consistent with the MWCNN, but the numbers of convolution blocks in the highest layer of two networks are different. In the MWCNN, there are eight convolution blocks in the highest layer, while in our MWRDCNN there are five convolution blocks in the highest layer (the last one is used to ensure the symmetry of the network). Therefore, the number of convolution blocks in the highest level of the MWDDCNN is three less than that of the MWCNN, which can accelerate the computation to a certain extent.

\section{Experiments}
In this section, we first show the training sets and training procedure for each network, then we make a fair comparison to demonstrate the performance of our network.

\subsection{Experimental Setting}
\subsubsection{Training and Test Sets}
Here, the DIV2K (one of the most popular training sets since 2017 \cite{PLiu-IA-2019,CLiu-IA-2019}, benefiting from its aesthetically high quality) is used for our training procedure \cite{Agustsson-CVPRW-2017}. It contains 800 images for training, 100 images for validation, and 100 images for testing.

In this work, we chose six other different networks as our main competitors. The networks and corresponding training patches and batch sizes were shown in Table~\ref{table1}.
\begin{table}[ht]
\renewcommand{\arraystretch}{1.3}
\caption{Patch and batch sizes for training different networks}
\label{table1}
\centering
\begin{tabular}{ccc}
\hline
Methods&Patch size&Batch size\\
\hline
IRCNN \cite{Zhang-CVPR-2017} & $152\times152$ & 32\\
DnCNN \cite{Zhang-TIP-2017} & $152\times152$ & 32\\
RED30 \cite{Mao-ANIPS-2016} & $152\times152$ & 21\\
MWCNN \cite{PLiu-IA-2019} & $152\times152$ & 32\\
RDN \cite{Zhang-CVPR-2018} & $76\times76$ & 28\\
MemNet \cite{Tai-ICCV-2017} & $76\times76$ & 18 \\
MWRDCNN & $152\times152$ & 32\\
\hline
\end{tabular}
\end{table}

Due to the limited GPU memory size, we set the batch size to 21 for the RED30. To adjust the batch size into a suitable value, we cropped every $152\times152$ patch to four $76\times76$ patches for training both the RDN and MemNet. By this means, we ensured that each network learned the same features from the training set.

The main task of trained networks is to denoise the images with additive Gaussian noise of different standard deviations. In this paper, we took three standard deviations of additive Gaussian noise into account, $\sigma=15$, 25 and 50. The performance of these networks were assessed by using five commonly used test sets: Set5 \cite{Bevilacqua-BMVC-2012}, Set12 \cite{Zhang-TIP-2017}, Set14 \cite{Roman-ICCS-2012}, BSD68 \cite{Martin-ICCV-2001} and Urban100 \cite{Huang-CVPR-CVPR}.

\subsubsection{Network Training}
To compare the proposed method with the previous networks as fair as possible, we adopted the same parameters when training different networks. Adaptive moment (Adam) estimation algorithm was chosen as the optimizer, which was set to $\alpha=0.01$, $\beta_1=0.9$, $\beta_2=0.999$ and $\epsilon=10^{-8}$. The learning rates were divided into 3 stages of 45 epochs in total. In the first 15 epochs, the learning rates were fixed to $10^{-3}$. In the next 20 epochs, the learning rates decayed exponentially from $10^{-3.8}$ to $10^{-4}$. In the last 10 epochs, the learning rates decayed exponentially from $10^{-4.5}$ to $10^{-5}$. The patches rotation and flip were used for data augmentation. We accomplished the training and test procedure on a NVIDIA RTX 2080Ti GPU with a package named MatConvNet \cite{Vedaldi-ACM-2015}.

\subsection{Comparisons with State-of-art Networks}
The performance of the denoised images and running time were evaluated in this section. The PSNR and structural similarity index measure (SSIM) \cite{Wang-TIP-2004} were utilized in our quantitative assessment. In this work, the images used in both training set and test sets were of gray-scale.

\subsubsection{Quantitative Evaluation}
The results of the proposed method and six previous methods on different Gaussian noise level were shown in Table~\ref{table2}. The highest score in each row is highlighted in bold. It can be seen that except the MemNet slightly exceeds ours in Urban100 test set, our MWRDCNN performs better than the other methods. It is worth noting that our MWRDCNN surpasses the MWCNN (the basis of our network) about 0.1~dB in the first four test sets, and about 0.3~dB in Urban100 test set in terms of PSNR. These results indicate that the proposed network does benefit from the residual learning from the RDBs. Particularly, all the networks with RDBs (the RDN, MemNet and MWRDCNN) perform much better than the others in Urban100 test set. Since many images from this test set contains various grids, the RDB may be gifted to deal with this kind of complicated patterns. On the other hand, the MWCNN gets a higher score than the RDN and MemNet in Set5, which is composed by many creature images. It can be seen from this result that the DWT and iDWT may help the network enlarge the RF and improve the fitting ability. Therefore, our MWRDCNN inherits this property and further improves the performance. In a nutshell, Our MWRDCNN combines the advantages of both the MWCNN and RDB together, and can competent in denoising the images of various types.
\begin{table*}[ht]
\renewcommand{\arraystretch}{1.3}
\caption{Average PSNR (dB) and SSIM values of our approach and six state-of-the-art methods in image denoising tasks.}
\label{table2}
\centering
\begin{tabular}{ccccccccc}
\hline
Test set&$\sigma$&IRCNN \cite{Zhang-CVPR-2017}&DnCNN \cite{Zhang-TIP-2017}&RED30 \cite{Mao-ANIPS-2016}&MWCNN \cite{PLiu-IA-2019}&RDN(G32C6D10) \cite{Zhang-CVPR-2018}&MemNet \cite{Tai-ICCV-2017}&MWRDCNN\\
\hline
& 15 & 33.48 / 0.9075 & 33.54 / 0.9087 &33.69 / 0.9047& 33.73 / 0.9063&-&-& $\mathbf{33.76 / 0.9067}$\\
Set5 & 25 & 31.17 / 0.8696 & 31.23 / 0.8709 & 31.43 / 0.8655 & 31.47 / 0.8680 &31.47 / 0.8773 &-& $\mathbf{31.55 / 0.8694}$\\
& 50 & 28.08 / 0.8009 & 28.13 / 0.8026 & 28.51 / 0.7999 & 28.63 / 0.8055 & 28.46 / 0.8152 & 28.49 / 0.8172 & $\mathbf{28.70 / 0.8079}$\\
\hline
& 15 & 32.73 / 0.9008 & 32.80 / 0.9019 & 32.97 / 0.9052&33.00 / 0.9065&-&-& $\mathbf{33.06 / 0.9075}$\\
Set12 &25&30.31 / 0.8589&30.38 / 0.8602&30.60 / 0.8663&30.61 / 0.8678&30.67 / 0.8681&-& $\mathbf{30.71 / 0.8696}$\\
&50&27.09 / 0.7799&27.14 / 0.7816&27.44 / 0.7938&27.56 / 0.7990&27.55 / 0.7977&27.62 / 0.7999& $\mathbf{27.66 / 0.8023}$\\
\hline
& 15 &32.31 / 0.8880&32.38 / 0.8891&32.56 / 0.8922&32.58 / 0.8939&-&-& $\mathbf{32.63 / 0.8945}$\\
Set14 &25&29.91 / 0.8354&29.96 / 0.8363&30.20 / 0.8422&30.22 / 0.8441&30.26 / 0.8439&-& $\mathbf{30.30 / 0.8463}$\\
&50&26.78 / 0.7359&26.80 / 0.7366&27.11 / 0.7514&27.21 / 0.7579&27.21 / 0.7547&27.26 / 0.7557& $\mathbf{27.32 / 0.7613}$\\
\hline
 & 15&31.62 / 0.8888&31.66 / 0.8894&31.78 / 0.8925&31.80 / 0.8935&-&-& $\mathbf{31.83 / 0.8940}$\\
BSD68&25&29.12 / 0.8244&29.16 / 0.8255&29.30 / 0.8313&29.33 / 0.8331&29.35 / 0.8325&-& $\mathbf{29.38 / 0.8351}$\\
&50&26.16 / 0.7156&26.18 / 0.7171&26.36 / 0.7263&26.44 / 0.7327&26.43 / 0.7320&26.45 / 0.7327& $\mathbf{26.50 / 0.7355}$\\
\hline
 & 15&32.42 / 0.9245&32.61 / 0.9266&33.05 / 0.9323&32.99 / 0.9334&-&-& $\mathbf{33.09 / 0.9345}$\\
Urban100&25&29.71 / 0.8814&29.88 / 0.8835&30.45 / 0.8953&30.38 / 0.8973&30.62 / 0.8983&-& $\mathbf{30.62 / 0.9010}$\\
&50&26.09 / 0.7885&26.18 / 0.7900&26.92 / 0.8157&27.04 / 0.8252&27.21 / 0.8252& $\mathbf{27.40 / 0.8311}$ & 27.36 / 0.8333\\
\hline
\end{tabular}
\end{table*}

\subsubsection{Qualitative Comparison}
\begin{figure*}[htb]
\centering
\includegraphics[scale=0.6]{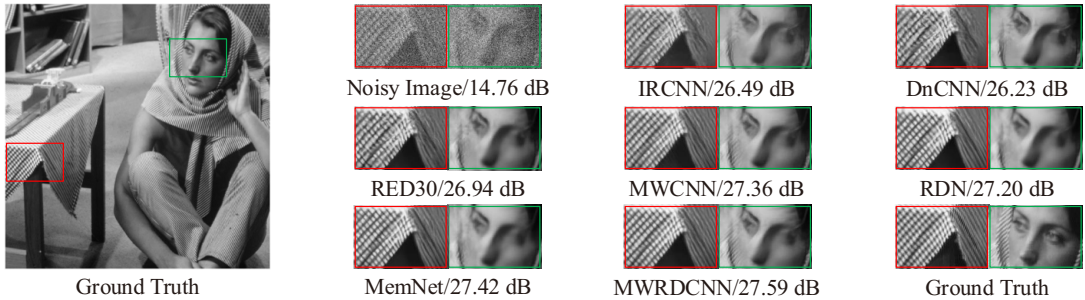}
\caption{Qualitative comparison using ``09" in Set12 with the standard deviation of the additive Gaussian noise $\sigma=50$.}
\label{fig:barbara}
\end{figure*}

\begin{figure*}[htb]
\centering
\includegraphics[scale=0.6]{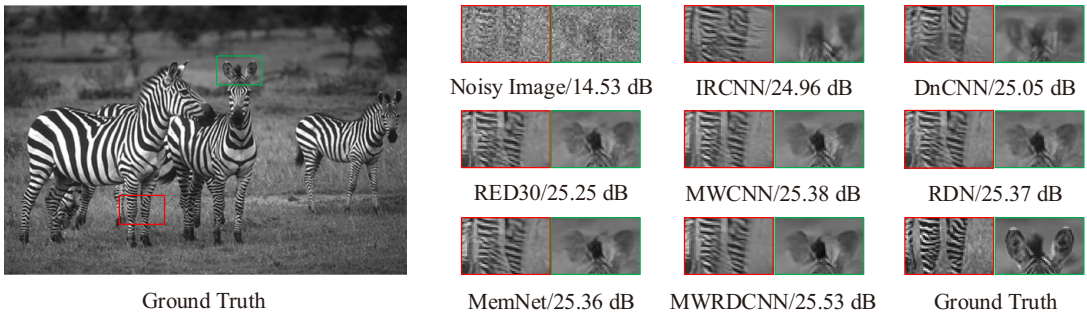}
\caption{Qualitative comparison using ``test051" in BSD68 with the standard deviation of the additive Gaussian noise $\sigma=50$.}
\label{fig:zebra}
\end{figure*}

We selected ``09" in Set12 and ``test051" in BSD68 for qualitatively comparing our method with the others. In each image, we enlarged two cropped patches for comparison. The PSNR values of the whole images obtained by different methods are listed under the corresponding patches. As shown in Fig.~\ref{fig:barbara}, the first three methods cannot correctly restore the tablecloth texture in the red rectangles. The MWCNN, RDN and MemNet can roughly recover the image, but there still exists some tiny flaws on the cloth edge. The result obtained by the proposed method has the highest similarity with the ground truth (original image without noise). In the green rectangles, our MWRDCNN obtained the clearest details, such as the right eye of the woman and the kerchief. The results of images with various texture directions are given in Fig.~\ref{fig:zebra}. Although all methods will reconstruct some redundant stripes beyond the zebra's legs, this negative phenomenon is not very significant in our method.

Overall, the MWRDCNN is promising in concisely recovering images which are badly interrupted by heavy noise and presents a better image denoising performance.

\subsubsection{Running Time}
As the network architectures become more and more complicated nowadays, the computational efficiency must be taken into account. In this work, we timed the mentioned methods in the same test environment. We adopted the CuDNN-v7.5 deep learning library with CUDA 10.1 to build the environment under Windows 10 operating system. After finishing the computational process with a NVIDIA RTX 2080ti GPU, we obtained the average time cost, as shown in Table~\ref{table3}.
\begin{table}[ht]
\renewcommand{\arraystretch}{1.3}
\caption{Running time (s) of different networks in image denoising tasks. The standard deviation of the additive Gaussian noise is set to $\sigma=50$}
\label{table3}
\centering	
\begin{tabular}{cccccccc}
\hline
Methods & Set5 & Set12 & Set14 & BSD68 & Urban100\\
\hline
IRCNN \cite{Zhang-CVPR-2017} & 0.0017 & 0.0017 & 0.0018 & 0.0017 & 0.0276\\
DnCNN \cite{Zhang-TIP-2017} & 0.0041 & 0.0041 & 0.0042 & 0.0041 & 0.0430\\
RED30 \cite{Mao-ANIPS-2016} & 0.0360 & 0.0345 & 0.0652 & 0.0364 & 0.2146\\
MWCNN \cite{PLiu-IA-2019} & 0.0240 & 0.0235 & 0.0394 & 0.0227 & 0.1248\\
RDN \cite{Zhang-CVPR-2018} & 0.0875 & 0.0827 & 0.1746 & 0.0868 & 0.8788\\
MemNet \cite{Tai-ICCV-2017} & 0.2127 & 0.2032 & 0.3886 & 0.2409 & 1.4648\\
MWRDCNN & 0.0649 & 0.0447 & 0.0993 & 0.0478 & 0.3191\\
\hline
\end{tabular}
\end{table}

Benefit from the DWT, which can reduces the size of the feature maps, our MWRDCNN runs faster than other two RDB-structured networks, i.e., the RDN and MemNet. Although slower than the IRCNN, DnCNN and MWCNN, the high-quality reconstruction is sufficient to compensate the time cost. Thus, MWRDCNN achieve a good balance between the performance and efficiency.

\section{Conclusion}
In this work, we chose the MWCNN as the backbone of our network, and inserted the RDBs in each downsampling and upsampling procedure to build our MWRDCNN. The proposed network inherits the large RF and lossless information operations in the MWCNN by using the DWT and iDWT. In addition, the RDBs can help the network establish a dense short-term residual learning. The experiment results show that by combining these advantages, the proposed network achieves higher qualities in image denoising tasks. Moreover, our MWRDCNN takes shorter computation time compared with the RDB-adopted networks like the RDN and MemNet. Thus, MWRDCNN is a successful attempt to utilize the superiority of multiple networks. We hope our MWRDCNN can be further extended to other image denoising tasks, such as the SISR and image artifacts removal.

\ifCLASSOPTIONcompsoc
  \section*{Acknowledgments}
  This work was supported by the National Natural Science Foundation of China (Grant No. 61801022), the National Key Research and Development Program of China (Grant No. 2016YFE0131500), the Civil Space Project of China (Grant No. D040301), the International Science and Technology Cooperation Special Project of Beijing Institute of Technology (Grant No. GZ2018185101).
\else
  \section*{Acknowledgment}
\fi

\ifCLASSOPTIONcaptionsoff
  \newpage
\fi
\newpage
\bibliographystyle{IEEEtran}
\bibliography{IEEEabrv,ref}

\iftrue
\begin{IEEEbiography}
[{\includegraphics[width=1in,height=1.25in]{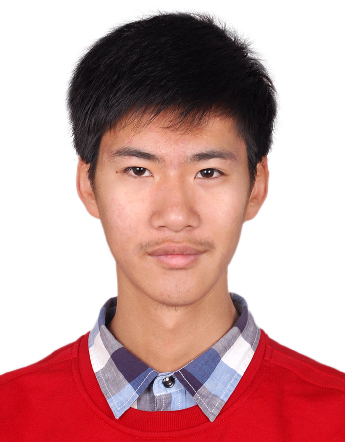}}]
{Shuo-Fei Wang}
received the BSc degree in physics from Beijing Institute of Technology, China, in 2018. He is currently working towards the PhD degree also in this university. His research interests include machine learning, image denoising, inverse problems in computer vision, and computational imaging.
\end{IEEEbiography}
\vspace{-13 cm}
\begin{IEEEbiography}
[{\includegraphics[width=1in,height=1.25in]{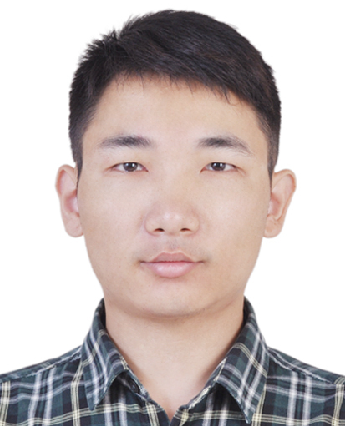}}]
{Wen-Ka Yu}
received the PhD degree in computer applied technology from University of Chinese Academy of Sciences, China, in 2015. He is currently an associate research professor with the School of Physics, Beijing Institute of Technology, China. His research interests include computer vision, compressed sensing, computational imaging, image reconstruction, image processing, image denoising, machine learning, and intelligent signal acquisition.
\end{IEEEbiography}
\vspace{-13 cm}
\begin{IEEEbiography}
[{\includegraphics[width=1in,height=1.25in,clip,keepaspectratio]{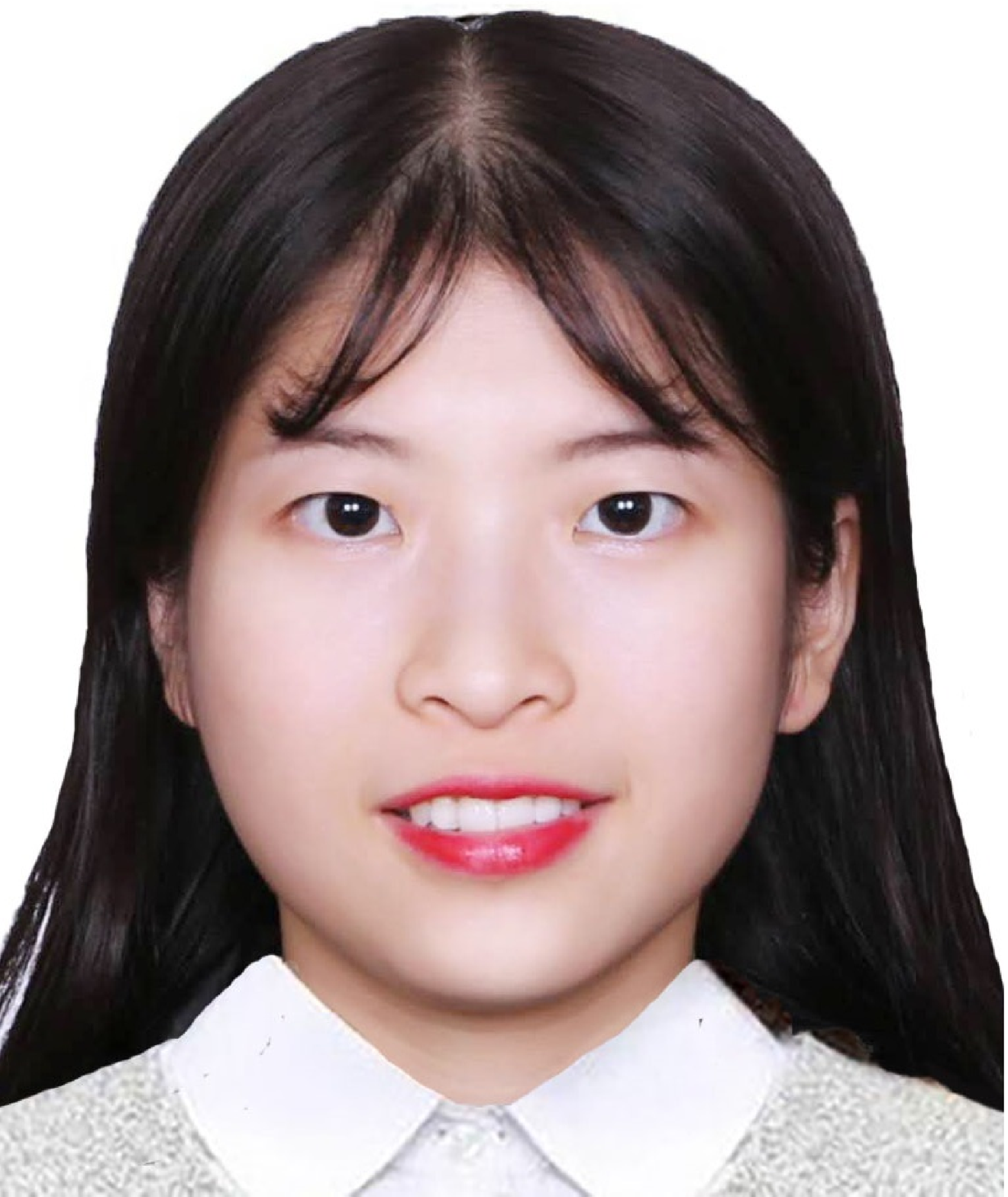}}]
{Ya-Xin Li}
received the BSc degree in physics from Northeast Normal University, China, in 2018. She is currently working towards the MSc degree in School of Physics, Beijing Institute of Technology, China. Her research interests include single-pixel imaging, image processing, image reconstruction, image denoising, and optical communication.
\end{IEEEbiography}

\noindent$\triangleright$ \textbf{For more information on this or any other computing topic, please visit our Digital Library.}

\end{document}